\documentclass[a4paper,11pt]{article}

\usepackage{amsmath,amsfonts,amsthm,amssymb}

\newtheorem{Claim}{Claim}
\newtheorem{Lemma}{Lemma}
\newtheorem{Theorem}{Theorem}
\newtheorem{Definition}{Definition}
\newtheorem{Corollary}{Corollary}
\newtheorem{Remark}{Remark}
\newcommand{\comment}[1]{}

\def\proof{\noindent{\bf Proof~}}

\begin{document}
\title{New upper bound on block sensitivity and certificate complexity in terms of sensitivity}
\author{Andris Ambainis \and Yihan Gao \and Jieming Mao \and Xiaoming Sun \and Song Zuo}
\date{}
\maketitle

\abstract{
Sensitivity ~\cite{CD82,CDR86} and block sensitivity ~\cite{Nisan91} are two important complexity measures of Boolean functions. A longstanding open problem in decision tree complexity, the ``Sensitivity versus Block Sensitivity'' question, proposed by Nisan and Szegedy~\cite{Nisan94} in 1992, is whether these two complexity measures are polynomially related, i.e., whether $bs(f)=O(s(f)^{O(1)})$.

We prove an new upper bound on block sensitivity in terms of sensitivity: $bs(f) \leq 2^{s(f)-1} s(f)$. Previously, the best upper bound on block sensitivity was $bs(f) \leq (\frac{e}{\sqrt{2\pi}}) e^{s(f)} \sqrt{s(f)}$ by Kenyon and Kutin~\cite{KK}.
We also prove that if $\min\{s_0(f),s_1(f)\}$ is a constant, then sensitivity and block sensitivity are linearly related, i.e. $bs(f)=O(s(f))$.
}

\section{Introduction}

%There is a large class of complexity measures for Boolean functions that study the low-level complexity for functions. This class of measures includes deterministic query complexity (also called decision tree complexity), sensitivity, block sensitivity, certificate and degree complexity. There is a long line of research focused on obtaining upper or lower bound of one complexity measure in terms of the others. We say that two complexity measures M and N are polynomially related if there are two constants $c_1$ and $c_2$ such that for any function $f$, $M(f) = O(N(f)^{c_1})$ and $N(f) = O(M(f)^{c_2})$. It is known that all complexity measures mentioned above, with the notable exception of sensitivity, are polynomially related with each other. The question whether sensitivity is polynomially related with other complexity measures, first asked by Nisan and Szegedy~\cite{Nisan94}, remains an open question in the study of Boolean functions.

Sensitivity ~\cite{CD82,CDR86} and block sensitivity ~\cite{Nisan91} are two important complexity measures in decision tree complexity. It has been proven that block sensitivity is polynomially related to a number of other complexity measures: certificate complexity, polynomial degree, approximate degree, deterministic/randomized/quantum query complexity, etc. \cite{BW02}. However, it is not known whether the simpler-to-define sensitivity has a similar property. The {\em sensitivity-block sensitivity} conjecture proposed by Nisan and Szegedy~\cite{Nisan94} suggests that for any total Boolean function $f$, the block sensitivity complexity and sensitivity complexity are polynomially related, i.e. $bs(f)=O(s(f)^{O(1)})$.

%For any Boolean function $f$, the sensitivity, block sensitivity, certificate complexity are denoted by $s(f)$, $bs(f)$ and $C(f)$, respectively. We also introduce another complexity measure, consistent block sensitivity, denoted by $cbs(f)$. The complexity measures mentioned above are all defined as the maximum value of certain evaluation over all input, therefore we also introduce the corresponding one side complexity measure, namely the maximum value over all input with the same function value. For any such complexity measure $M$, the one side complexity measure of $M$ are denoted by $M_0(f)$ and $M_1(f)$ respectively. The definition of these complexity measures (and their one side version) will be given in section~\ref{preliminary}.

The best known lower bound on $bs(f)$ in terms of $s(f)$ is quadratic and upper bound is exponential. Rubinstein~\cite{Rubinstein95} gave the first quadratic separation for block sensitivity and sensitivity by constructing a Boolean function $f$ with $bs(f)={1\over 2}s(f)^2$. Virza~\cite{Virza10} improved the separation to $bs(f) =
{1\over 2}s(f)^2 + s(f)$. Ambainis and Sun~\cite{AS11} further improved the gap to $bs(f) = {2\over 3}s(f)^2-{1\over 2}s(f)$.
 However, the best known upper bound on block sensitivity is
\begin{equation}
bs(f) \leq (\frac{e}{\sqrt{2\pi}}) e^{s(f)} \sqrt{s(f)},
\end{equation}
given by Kenyon and Kutin \cite{KK}. More information about sensitivity-block sensitivity conjecture can be found in a survey paper by Hatami et al.~\cite{HKP11} and in Scott Aaronson's blog post \cite{Aaronson10}.

In this paper, we give a new upper bound on certificate complexity (which is related to block sensitivity).

\begin{Theorem}\label{thm:C(f)}
 For any Boolean function $f$,
\begin{equation}
C_1(f)\leq 2^{s_0(f)-1} s_1(f),\ \ \
C_0(f)\leq 2^{s_1(f)-1} s_0(f).
\end{equation}
\end{Theorem}
This implies
\begin{Corollary}
\label{cor:1}
\begin{equation*}
bs(f)\leq C(f) \leq 2^{s(f)-1} s(f).
\end{equation*}
\end{Corollary}

We also introduce a new complexity measure called consistent block sensitivity, see section~\ref{sec:thm2} for the definition. Using it we prove another upper bound for block sensitivity.
\begin{Theorem}\label{thm:bs(f)}
For any Boolean function $f$,
\begin{equation}
bs(f)\leq \min\{2^{s_0(f)},2^{s_1(f)}\} s_1(f)s_0(f).
\end{equation}
\end{Theorem}
One consequence of this result is that,
\begin{Corollary}
If $\min\{s_0(f),\ s_1(f)\}=O(1)$, then
$bs(f)=O(s(f))$.
\end{Corollary}
\begin{Remark}
This result suggests a stronger version of sensitivity-block sensitivity conjecture: %is it true that
\begin{equation}
bs(f)=O(s_1(f)s_0(f))?
\end{equation}
\end{Remark}

The rest part of the paper is organized as follows: in Section~\ref{sec:preliminary} we give some basic definitions. We prove Theorem~\ref{thm:C(f)} in Section~\ref{sec:thm1} and Theorem~\ref{thm:bs(f)} in section~\ref{sec:thm2}. We conclude the paper in Section~\ref{sec:con}.

%In Section 3, we consider consistent block sensitivity and show $bs_0(f) \leq ( C_1(f) - \frac{1}{2}) s_0(f)$. In Section 2, we give an upper bound on $C(f)$: $C_1(f)\leq 2^{s_0(f) - 1} s_1(f) - s_0(f) + 1$. Combining them, we show that $bs_0(f) \leq \min\{(2^{s_0(f)-1} s_1(f)-s_0(f) + 1)s_0(f),2^{s_1(f) - 1} s_0(f) - s_1(f) + 1\}$.
%Previously, the best upper bound on $bs(f)$ was
%$bs(f) \leq (\frac{e}{\sqrt{2\pi}}) e^{s(f)} \sqrt{s(f)}$ by Kenyon and Kutin \cite{KK}.

\section{Preliminaries}\label{sec:preliminary}

In this section, we give the basic definitions we use in this paper. %of sensitivity, block sensitivity and certificate complexity.
See~\cite{BW02} for a survey on more complexity measures for decision trees and the relationships between them.
Let $f:\{0,1\}^n\rightarrow\{0,1\}$ be a total Boolean function.

\begin{Definition}
The {\em sensitivity complexity} $s(f,x)$ of $f$ on input $x$ is defined as the number of bits on which the function is sensitive, i.e. $s(f,x)=\big|\{i|f(x)\neq f(x^{(i)})\}\big|$, where $x^{(i)}$ is obtained by flipping the $i$-th bit of $x$. Define the {\em sensitivity} of $f$ as $s(f)=\max\big\{s(f,x)|x\in\{0,1\}^n\big\}$ and
the {\em 0-sensitivity} and {\em 1-sensitivity} of $f$ as $s_0(f)=\max\big\{s(f,x)|x\in\{0,1\}^n,f(x)=0\big\}$, $s_1(f)=\max\big\{s(f,x)|x\in\{0,1\}^n,f(x)=1\big\}$.
\end{Definition}

\begin{Definition}
The {\em block sensitivity complexity} $bs(f,x)$ of $f$ on input $x$ is defined as maximum number of pairwise disjoint subsets $B_1,...,B_k$ of $[n]$ such that $f(x) \neq f(x^{(B_i)})$ for all $i\in [k]$. Here $x^{(B_i)}$ is obtained by flipping all the bits $\{x_j|j\in B_i\}$ of $x$. Define the {\em block sensitivity} of $f$ as $bs(f)=\max\big\{bs(f,x)|x\in\{0,1\}^n\big\}$ and
the {\em 0-block sensitivity} and {\em 1-block sensitivity} of $f$ as $bs_0(f)=\max\big\{bs(f,x)|x\in\{0,1\}^n,f(x)=0\big\}$, $bs_1(f)=\max\big\{bs(f,x)|x\in\{0,1\}^n,f(x)=1\big\}$.
\end{Definition}

\begin{Definition}
The {\em certificate complexity} $C(f,x)$ of $f$ on input $x$ is defined as the minimum length of a partial assignment of $x$ such that $f$ is constant on this restriction. Define the {\em certificate complexity} of $f$ as $C(f)=\max\big\{C(f,x)|x\in\{0,1\}^n\big\}$ and
the {\em 0-certificate} and {\em 1-certificate} of $f$ as $C_0(f)=\max\big\{C(f,x)|x\in\{0,1\}^n,f(x)=0\big\}$, $C_1(f)=\max\big\{C(f,x)|x\in\{0,1\}^n,f(x)=1\big\}$.
\end{Definition}

We introduce a new notation called {\em consistent block sensitivity} here.
\begin{Definition}
The {\em consistent block sensitivity} of $f$ on an input $x$
(denoted $cbs(f,x)$) is equal to
the maximum number $k$ of blocks $B_1, \ldots, B_k$
such that $f(x)\neq f(x^{(B_i)})$ and $x^{(B_1)}, \ldots, x^{(B_k)}$
have certificates $C_1, \ldots, C_k$ such that there is an input $y$ that
satisfies all of $C_i$ simultaneously.
The {\em consistent block sensitivity} of $f$
(denoted $cbs(f)$) is equal to the maximum of $cbs(f,x)$ over all inputs $x$.
The $c$-{\em consistent block sensitivity} of $f$
(denoted $cbs_c(f)$) is equal to the maximum of $cbs(f,x)$ over all inputs $x$
such that $f(x)=c$.
\end{Definition}

\section{Upper bound for certificate complexity in terms of sensitivity}\label{sec:thm1}

In this section we prove theorem~\ref{thm:C(f)}. Actually, we prove a slightly stronger result.

\begin{Theorem}\label{thm:C(f)-2}
Let $f:\{0,1\}^n\rightarrow\{0,1\}$ be a Boolean function, then
\begin{equation}
C_1(f)\leq 2^{s_0(f) - 1} s_1(f) - (s_0(f) - 1),\ C_0(f)\leq 2^{s_1(f) - 1} s_0(f) - (s_1(f) - 1).\footnote{If $s_0(f)=0$ or $s_1(f)=0$, then $f$ is constant, hence $s(f)=bs(f)=C(f)=0$.}
\end{equation}
\end{Theorem}

\proof
By symmetry we only need to prove $C_1(f)\leq 2^{s_0(f) - 1} s_1(f) - (s_0(f) - 1)$.
 Without the loss of generality, we assume $C_1(f)=C(f,0^n)$, i.e. the
1-certificate complexity is achieved on the input $0^n$. We have $f(0^n)=1$. We assume that the minimal certificate of $0^n$
consists of $x_1=0, x_2=0,\ldots,x_m=0$, where $m=C(f,0^n)=C_1(f)$.

Let $Q_0$ be the set of inputs $x$ that satisfies $x_1 =x_2=\ldots=x_{m} = 0$. Since $x_1=0, x_2=0,\ldots,x_m=0$ is a 1-certificate, we have $\forall~x\in Q_0$, $f(x)=1$.

For each $i \in [m]$, let $Q_i$ be the set of inputs $x$ with $x_1 =\ldots=x_{i-1} = x_{i+1} = \ldots= x_m = 0$ and $x_i = 1$. Let $S$ be the total sensitivity of all inputs $x \in \bigcup_{i=1}^m Q_i$. It consists of three parts: sensitivity between $Q_i$ and $Q_0$~(denoted by $S_1$), sensitivity inside $Q_i$~(denoted by $S_2$) and sensitivity between $Q_i$ and other input~(denoted by $S_3$), i.e.
\begin{equation}
S = \sum_{i=1}^{m}\sum_{x \in Q_i} s(f,x) = S_1 + S_2 + S_3.
\end{equation}

In the following we estimate $S_1,S_2$ and $S_3$ separately. We use $A_1, \ldots, A_m$ to denote the set of $0$-inputs in $Q_1, \ldots, Q_m$, respectively, i.e. $A_i=\{x\in Q_i|f(x)=0\}$~($i\in [m]$). Since $x_1=\ldots=x_m=0$ is the minimal certificate, i.e. $Q_0$ is maximal, thus $A_1, \ldots, A_m$ are all nonempty.

We will use the following isoperimetric inequality for
the Boolean hypercube.
\begin{Lemma}\label{lem:iso}\cite{Bol}
For any $A\subseteq \{0,1\}^n$, the edges between $A$ and $\bar{A}=\{0,1\}^n\setminus A$ is lower bounded by
$$|E(A,\bar{A})|\geq |A|(n-\log_2 |A|).$$
\end{Lemma}
We also need the following lemma which follows from Lemma \ref{lem:iso} but can be
also proven without using it \cite{Sim}:
\begin{Lemma}\label{lem:Ai}\cite{Sim}
For any $i\in [m]$, $$|A_i| \geq 2^{n-m-s_0(f)+1}.$$
\end{Lemma}

%\begin{proof}(of Lemma~\ref{lem:Ai})
%Consider the edges between $A_i$ and $Q_i\setminus A_i$, by using the isoperimetric %inequality  on $Q_i$, we have
%$$|E(A_i,Q_i\setminus A_i)|\geq |A_i|(\log_2{|Q_i|}-\log_2 |A_i|) =|A_i|(n-m-\log_2 |A_i|).$$

%Hence, for some $x\in A_i$, there are at least $(n-m-\log |A_i|)$ ``neighbors" in $Q_i\setminus A_i$, i.e. $(n-m-\log |A_i|)$ different inputs $y\in Q_i\setminus A_i$, $x$ and $y$ differ by only one bit.
%Since $A_i$ are the 0-input in $Q_i$, thus $\forall~y\in Q_i\setminus A_i$, $f(y)=1$. Therefore, $x$ is sensitivity on at least $(n-m-\log |A_i|)$ variables in $\{x_{m+1},\ldots,x_n\}$ (because $x_1,\ldots,x_m$ are fixed for $Q_i$). Since $f(x)=0$, and $\forall~y\in Q_0$, $f(y)=1$, thus $x$ is also sensitivity on variable $x_i$. Hence, $s(f,x)\geq (n-m-\log|A_i|)+1$. But $s(f,x)$ is no more than $s_0(f)$, combine these two we get
%$|A_i|\geq 2^{n-m+1-s_0(f)}$.
%\qed
%\end{proof}

The sensitivity between $Q_i$ and $Q_0$ is $|A_i|$. By summing over all possible $i$ we get:
\begin{equation}\label{eq5.3}
S_1=\sum_{i=1}^m |A_i|.
\end{equation}

{\bf Sensitivity inside $Q_1, \ldots, Q_m$:}
By Lemma~\ref{lem:iso}, for each $i\in [m]$, the number of edges between $A_i$ and $Q_i\setminus A_i$ is bounded by:
\begin{eqnarray*}
|E(A_i, Q_i\setminus A_i)| \geq  |A_i|(\log_2 |Q_i| - \log_2 |A_i|)=|A_i|(n-m-\log_2 |A_i|).
\end{eqnarray*}
Therefore,
\begin{eqnarray}
S_2&=& 2\sum_{i=1}^m |E(A_i, Q_i\setminus A_i)|\nonumber\\
&\geq &2\sum_{i=1}^m |A_i|(n-m - \log_2|A_i|).\label{eq5.4}
\end{eqnarray}

{\bf Sensitivity between $Q_i$ and other inputs:}
For each $1 \leq i < j \leq m$, let $Q_{i,j}$ be the set of inputs that are {\em adjacent} to both $Q_i$ and $Q_j$, i.e. $Q_{i,j}$ is the set of inputs $x$ that satisfy $x_1 =\ldots x_{i-1}=x_{i+1}=\ldots x_{j-1}=x_{j+1}=\ldots x_m = 0$ and $x_i = x_j = 1$. The sensitivity between $Q_i, Q_j$ and $Q_{i,j}$ is lower bounded by
\begin{eqnarray*}
\sum_{x \in Q_0} |f(x+e_i) - f(x+e_j)|.
\end{eqnarray*}
where $e_i$ is the unit vector with the $i$-th coordinate equal to 1 and all other coordinates equal to 0. Then,
$x+e_i$, $x+e_j$ are the neighbors of $x$ in $Q_i$ and $Q_j$, respectively. Summing over all possible pairs of $(i,j)$ we get
\begin{eqnarray}
\nonumber
S_3&\geq& \sum_{1\leq i < j\leq m} \sum_{x\in Q_0} |f(x+e_i) - f(x+e_j)|\\
& =& \sum_{x\in Q_0} \left(\sum_{i=1}^m f(x+e_i)\right)\left(m - \sum_{i=1}^m f(x+e_i)\right)\nonumber\\
&=& \sum_{x\in Q_0} s(f,x)(m-s(f,x)).\label{eq5.5}
\end{eqnarray}

\vskip10pt
If we combine inequalities~(\ref{eq5.3})-(\ref{eq5.5}), we get
\begin{eqnarray}
S &=& \sum_{i=1}^{m}\sum_{x\in Q_i}s(f,x)\nonumber\\
 &\geq& \sum_{i=1}^m |A_i| + 2\sum_{i=1}^m |A_i|(n-m - \log_2|A_i|)+ \sum_{x\in Q_0} s(f,x)(m - s(f,x)).\label{eq5.6}
\end{eqnarray}
Since $s(f,x)$ is upper bounded by $s_0(f)$ or $s_1(f)$ (depending
on whether $f(x)=0$ or $f(x)=1$), we have
\begin{eqnarray*}
\sum_{x\in Q_i}s(f,x)&\leq& |A_i|s_0(f) + (|Q_i| - |A_i|)s_1(f)\\
&=&|A_i|s_0(f) + (2^{n-m} - |A_i|)s_1(f)
\end{eqnarray*}
Thus,
\begin{eqnarray}
S=\sum_{i=1}^{m}\sum_{x\in Q_i}s(f,x) \leq \sum_{i=1}^m \Big(|A_i|s_0(f) + (2^{n-m} - |A_i|)s_1(f)\Big).\label{eq7}
\end{eqnarray}
We use $w$ to denote the total number of $0$-inputs in $Q_1, \ldots, Q_m$. Then,
\begin{eqnarray*}
w = \sum_{i=1}^m |A_i| = \sum_{x \in Q_0} s(f,x).
\end{eqnarray*}
The inequality~(\ref{eq7}) can be rewritten as
\begin{equation}
S \leq  w\cdot s_0(f) + (m \cdot 2^{n-m} -w)s_1(f).\label{eq5.7}
\end{equation}
Also, $s(f,x)\leq s_1(f)$ for each $x\in Q_0$.
Thus, the right-hand side of inequality~(\ref{eq5.6}) is
\begin{eqnarray}
\nonumber
&&\sum_{i=1}^m |A_i| + 2\sum_{i=1}^m |A_i|(n-m - \log_2|A_i|) + \sum_{x\in Q_0} s(f,x)(m - s(f,x))\\
\nonumber
 &\geq& w + 2\sum_{i=1}^m |A_i|(n-m - \log_2|A_i|) + (m - s_1(f))\sum_{x\in Q_0} s(f,x)\\
\nonumber
&= & w + 2w(n-m)-2\sum_{i=1}^m |A_i|\log_2|A_i| + (m - s_1(f))w\\
&= & w(1+2n-m - s_1(f)) - 2\sum_{i=1}^m |A_i|\log_2|A_i|.\label{eq5.8}
\end{eqnarray}
By combining inequalities~(\ref{eq5.6})-(\ref{eq5.8}) we get
\begin{eqnarray*}
w(1+2n-m - s_1(f)) - 2\sum_{i=1}^m |A_i|\log_2|A_i|
\leq w\cdot s_0(f)+(m\cdot 2^{n-m}-w)s_1(f).%=(s_1 - s_0)S + 2^V N s_0
\end{eqnarray*}
By rearranging the inequality we get
\begin{equation}
w(1 + 2n-m  - s_0(f)) \leq 2\sum_{i=1}^m |A_i| \log_2 |A_i| + m\cdot 2^{n-m} s_1(f).\label{eq8}
\end{equation}

Since the function $g(x)=x \log_2 x$ is convex, we know that
$|A_i|\leq |Q_i|=2^{n-m}$. From Lemma~\ref{lem:Ai} $|A_i|\geq 2^{n-m-s_0(f)+1}$.
Therefore,
\begin{eqnarray*}
g(|A_i|)&=&g\left(\frac{|A_i|-2^{n-m-s_0(f)+1}}{2^{n-m}-2^{n-m-s_0(f)+1}}\cdot 2^{n-m}+\frac{2^{n-m}-|A_i|}{2^{n-m}-2^{n-m-s_0(f)+1}}\cdot 2^{n-m-s_0(f)+1}\right)\\
&\leq&
\frac{|A_i|-2^{n-m-s_0(f)+1}}{2^{n-m}-2^{n-m-s_0(f)+1}}\cdot g(2^{n-m})+\frac{2^{n-m}-|A_i|}{2^{n-m}-2^{n-m-s_0(f)+1}}\cdot g(2^{n-m-s_0(f)+1})\\
&=&\frac{|A_i|-2^{n-m-s_0(f)+1}}{2^{n-m}-2^{n-m-s_0(f)+1}}\cdot 2^{n-m}(n-m)\\
&& +\frac{2^{n-m}-|A_i|}{2^{n-m}-2^{n-m-s_0(f)+1}}\cdot 2^{n-m-s_0(f)+1}(n-m-s_0(f)+1)\\
&=&\frac{|A_i|-2^{n-m-s_0(f)+1}}{2^{s_0(f)-1}-1}\cdot 2^{s_0(f)-1}(n-m) +\frac{2^{n-m}-|A_i|}{2^{s_0(f)-1}-1}(n-m-s_0(f)+1)\\
&=&\left(\frac{|A_i|-2^{n-m-s_0(f)+1}}{2^{s_0(f)-1}-1} 2^{s_0(f)-1}+\frac{2^{n-m}-|A_i|}{2^{s_0(f)-1}-1}\right)(n-m) -\frac{2^{n-m}-|A_i|}{2^{s_0(f)-1}-1}(s_0(f)-1)\\
&=&|A_i|(n-m) -\frac{2^{n-m}-|A_i|}{2^{s_0(f)-1}-1}(s_0(f)-1).
\end{eqnarray*}
Hence
\begin{eqnarray}
\nonumber \sum_{i=1}^m |A_i| \log_2|A_i|&=&\sum_{i=1}^{m}g(|A_i|)\\
\nonumber & \leq & \sum_{i=1}^m \left(|A_i|(n-m) - \frac{2^{n-m} - |A_i|}{2^{s_0(f) - 1} - 1}(s_0(f) - 1)\right)\\
&=& w(n-m+\frac{s_0(f)-1}{2^{s_0(f)-1}-1}) - 2^{n-m}\frac{s_0(f) - 1}{2^{s_0(f) - 1} - 1}.
\label{eq9}
\end{eqnarray}
By combining inequalities~(\ref{eq8}) and (\ref{eq9}), we get
\begin{eqnarray*}
&& w(1 + 2n-m - s_0(f))\\
&& \leq 2\left(w(n-m+\frac{s_0(f)-1}{2^{s_0(f)-1}-1}) - 2^{n-m}\frac{s_0(f) - 1}{2^{s_0(f) - 1} - 1}\right)  + m\cdot 2^{n-m} s_1(f).
\end{eqnarray*}
It implies that
\begin{eqnarray*}
w\left(1 + m - s_0(f) - \frac{2(s_0(f) - 1)}{2^{s_0(f)-1} - 1}\right) \leq m\cdot 2^{n-m} \left(s_1(f) - \frac{2(s_0(f) - 1)}{2^{s_0(f)-1} - 1}\right),
\end{eqnarray*}
Substituting $w =\sum_{i=1}^{m} |A_i|\geq m \cdot 2^{n-m-s_0(f)+1}$, we get
\begin{eqnarray*}
m\cdot 2^{n-m-s_0(f)+1} \left(1 + m - s_0(f) - \frac{2(s_0(f) - 1)}{2^{s_0(f)-1} - 1}\right) \leq m\cdot 2^{n-m} \left(s_1(f) - \frac{2(s_0(f) - 1)}{2^{s_0(f)-1} - 1}\right),
\end{eqnarray*}
i.e.
\begin{eqnarray*}
1 + m - s_0(f) - \frac{2(s_0(f) - 1)}{2^{s_0(f)-1} - 1} \leq 2^{s_0(f)-1} \left(s_1(f) - \frac{2(s_0(f) - 1)}{2^{s_0(f)-1} - 1}\right),
\end{eqnarray*}
which implies
$$m \leq 2^{s_0(f) - 1} s_1(f) - s_0(f) + 1.$$
\qed

\section{Upper bound for block sensitivity in terms of sensitivity}\label{sec:thm2}

First we show that the consistent block sensitivity is no more than sensitivity.

\begin{Theorem}
\label{thm:cbs1}
$s_b(f)\geq cbs_b(f)$ for any $b\in\{0, 1\}$.
\end{Theorem}

\proof We prove $s_0(f)\geq cbs_0(f)$ here.
Without the loss of generality assume that the maximum $cbs_0(f)=k$ is
achieved on the input $0^n$. Let $|B_i|=b_i$. We denote the variables
in the block $B_i$ as $x_{i, 1}, x_{i, 2}$, $\ldots$, $x_{i, b_i}$.
We further assume that blocks $B_i$ are minimal.

Since $C_i$ is a certificate for $x^{(B_i)}$, it must include conditions
$x_{i, 1}=1$, $x_{i, 2}=1$, $\ldots$, $x_{i, b_i}=1$ and no other conditions $x_j=1$.
(If $C_i$ included some other $x_j=1$, it would not be satisfied by $x^{(B_i)}$.
If $C_i$ did not include $x_{i, j}=1$, it would also be satisfied by
$x^{(B_i - (i, j))}$ - contradicting the assumption that $B_i$ is minimal.)

Moreover, $C_i$ must not include a condition $x_{i', j}=0$ for $i'\neq i$.
(Otherwise, it would be impossible to satisfy $C_i$ and $C_{i'}$ simultaneously.)
Hence, $C_i$ consists of $x_{i, 1}=1$, $x_{i, 2}=1$, $\ldots$, $x_{i, b_i}=1$
and, possibly, $x_j=0$ for some variables $j$ that do not belong to any of $B_i$.

\comment{We consider the function $f'$ obtained from $f$ by fixing $x_j=0$ for all
$j\notin B_1 \cup B_2 \cup \ldots \cup B_k$. After this substitution,
each $C_i$ consists of We claim}

We say that an input $y$ has a type $(t_1, \ldots, t_k)$
if $y_i=0$ for all $i\notin B_1\cup \ldots \cup B_k$ and,
for every $i\in\{1, \ldots, k\}$,
the number of $y_{i, j}=1$ is equal to $t_i$.

\begin{Claim}
\label{cl:contra}
Assume that $s_0(f)<k$. Then, for every input $y$ of
type $(t_1, \ldots, t_k)$, $t_1<b_1$, $\ldots$, $t_k<b_k$,
$f(y)=1$.
\end{Claim}

Substituting $t_1=\ldots=t_k=0$ leads to the conclusion $f(0^n)=1$
which is in a contradiction with our initial assumption $f(0^n)=0$.
Hence, the assumption $s_0(f)<k$ is wrong and we must have
$s_0(f)\geq k = cbs_0(f)$.

\proof (of Claim \ref{cl:contra})
By induction on $B=(b_1+\ldots+b_k)-(t_1+\ldots+t_k)$.

The smallest possible value is $B=k$ (if $t_i=b_i-1$ for all $i$).
This is our base case. In this case, there is one
variable $y_{i, j}=0$ for each $i$. Then $y^{(i, j)}$ satisfies the
certificate $C_i$. Hence, $f(y^{(i, j)})=1$.
This means that, if $f(y)=0$, then $y$ is sensitive to $k$ variables $y_{i, j}$ -
one for each $i\in\{1, \ldots, k\}$.
Thus, $s_0(f)<k$ implies $f(y)=1$.

For the inductive case, we assume that Claim is true for $B=B_0-1$ and prove that this
implies Claim being true for $B=B_0$.
Let $y$ be an input for which $B=B_0$.
We choose $y_{i, j}=0$ for each $i\in\{1, \ldots, k\}$.
(Such $y_{i, j}$ always exists, since $t_i<b_i$.)
We claim that $f(y^{(i, j)})=1$. (If $t_i=b_i-1$, then, with $y_{i, j}=1$
we have $y_{i, 1}=\ldots=y_{i, b_i}=1$ and the certificate $C_i$ is satisfied.
Hence, $f(y^{(i, j)})=1$. If $t_i<b_i-1$, then $y^{(i, j)}$ is
an input of type $(t_1, \ldots, t_i+1, \ldots, t_k)$ and
$f(y^{(i, j)})=1$ follows from the inductive assumption.) By the same argument
as in the previous case, this implies $f(y)=1$.
\qed

Next we upper bound block sensitivity by certificate complexity and consistent block sensitivity.
\begin{Theorem}
\label{thm:cbs2}
$bs_0(f) \leq  2 (C_1(f) - \frac{1}{2}) cbs_0(f),\ bs_1(f) \leq  2 (C_0(f) - \frac{1}{2}) cbs_1(f)$.
\end{Theorem}

Together with Theorem \ref{thm:cbs1}, this implies

\begin{Corollary}
\label{cor:cbs2}
$bs_0(f) \leq 2 ( C_1(f) - \frac{1}{2}) s_0(f),\ bs_1(f) \leq 2 ( C_0(f) - \frac{1}{2}) s_1(f)$.
\end{Corollary}

\proof (of Theorem \ref{thm:cbs2})
We use the same notation as in the proof of Theorem \ref{thm:cbs1}.
Also, we again assume that the maximum $bs_0(f)$ is
achieved on the input $x=0^n$ and the sensitive blocks $B_i$ are minimal.

We consider a graph $G$ with $m=bs_0(f)$ vertices $v_1$, $\ldots$, $v_{bs_0(f)}$,
with an edge $(v_i, v_j)$ if the certificates $C_i$ and $C_j$ are consistent.
The size of the largest clique in this graph is at most $cbs_0(f)$.
By Turan's theorem, a graph not containing a clique of size $k+1$, $k = cbs_0(f)$
has at most $\frac{k-1}{2k} m^2$ edges.

Therefore $\overline{G}$ (the complement of graph $G$) has at least
\begin{equation}
\label{eq:turan}
{m \choose 2} - \frac{k-1}{2k} m^2 = \frac{m^2}{2k} - \frac{m}{2}
\end{equation}
edges. Each of those edges corresponds to a pair of certificates $C_i$, $C_j$
that are not consistent.

Since $C_i$ is a certificate for $x^{(B_i)}$ and $B_i$ is minimal,
$C_i$ must consist of conditions $x_{i, 1}=1$, $\ldots$, $x_{i, b_i}=1$
and $x_j=0$ for some $j\notin B_i$. (This follows in the same way as
the similar statement in the proof of Theorem \ref{thm:cbs1}.)

If $C_i$ and $C_j$ are inconsistent, then either $C_i$ contains a condition $x_{j, l}=0$
(which is not consistent with $x_{j, l}=1$ in $C_j$) or $C_j$ contains a condition
$x_{i, l}=0$ (which is not consistent with $x_{i, l}=1$ in $C_i$).
Since $|C_i|\leq C_1(f)$, $C_i$ may contain conditions $x_{j, l}=0$ for at most $C_1(f)-1$
different $j$. This means that the number of edges in $\overline{G}$ is at most
$m(C_1(f)-1)$. Together with (\ref{eq:turan}), this implies
\[ \frac{m^2}{2k} - \frac{m}{2} \leq m(C_1(f)-1) , \]
\[ \frac{m}{2k} - \frac{1}{2} \leq (C_1(f)-1) , \]
\[ m \leq 2k \left( C_1(f) - \frac{1}{2} \right) .\]
\qed

Now we prove Theorem~\ref{thm:bs(f)}.

\proof (of Theorem \ref{thm:bs(f)}) From Theorem~\ref{thm:C(f)} $C_0(f)\leq 2^{s_1(f)-1}s_0(f)$, hence $$bs_0(f)\leq C_0(f)\leq 2^{s_1(f)-1}s_0(f).$$ From Corollary~\ref{cor:cbs2} we have $bs_0(f)\leq 2(C_1(f) - \frac{1}{2}) s_0(f)$, together with Theorem~\ref{thm:C(f)} we get
$$bs_0(f)\leq 2 (2^{s_0(f)-1}s_1(f)-\frac{1}{2}) s_0(f).$$
Therefore,
$$bs_0(f)\leq \min\{2^{s_1(f)}s_0(f)s_1(f),\ 2^{s_0(f)}s_1(f)s_0(f)\}.$$
Similarly we can show that
$$bs_1(f)\leq \min\{2^{s_1(f)}s_0(f)s_1(f),\ 2^{s_0(f)}s_1(f)s_0(f)\}.$$
\qed

\section{Conclusions}\label{sec:con}

In this paper we proved an upper bound for certificate complexity in terms of sensitivity, which also implies upper bound on block sensitivity in terms of sensitivity. We also introduce the concept of consistent block sensitivity. We then use it to prove that block sensitivity and sensitivity are linear related if $\min\{s_0(f),s_1(f)\}=O(1)$. This suggests a refined version of the sensitivity-block sensitivity question:
is it true that $$bs(f) = O(s_0(f)s_1(f))?$$


\begin{thebibliography}{9}


\bibitem{Aaronson10} S. Aaronson. My philomath project: Sensitivity versus block-sensitivity, http://www.scottaaronson.com/blog/?p=453, June 13, 2010.

\bibitem{AS11} Andris Ambainis and Xiaoming Sun. New separation between $s(f)$ and $bs(f)$. Electronic Colloquium on Computational Complexity (ECCC) 18: 116 (2011).

\bibitem{BW02} H. Buhrman and R. de Wolf. Complexity measures and decision tree
complexity: a survey, \textit{Theoretical Computer Science}
288(1): 21-43, 2002.

\bibitem{CD82}
Stephen Cook and Cynthia Dwork. Bounds on the Time for Parallel RAM's to Compute Simple
Functions. In {\em STOC}, pages 231-233, 1982.

\bibitem{CDR86}
Stephen Cook, Cynthia Dwork, and R$\ddot{u}$diger Reischuk. Upper and Lower Time Bounds for Parallel
Random Access Machines without Simultaneous Writes.
{\em SIAM J. Comput.,} 15(1):87-97, 1986.

\bibitem{Bol}
B. Bollobas. {\em Combinatorics: set systems,
hypergraphs, families of vectors and combinatorial probability.}
Cambridge University Press, Cambridge, 1986.

\bibitem{HKP11} P. Hatami, R. Kulkarni, D. Pankratov. Variations on the Sensitivity Conjecture,
\textit{Theory of Computing Library, Graduate Surveys} No. 4 (2011) pp. 1-27.

\bibitem{KK}
C. Kenyon, S. Kutin. Sensitivity, block sensitivity, and l-block sensitivity of
Boolean functions. {\em Information and Computation}, 189(1):43, 2004.

\bibitem{Nisan91} N. Nisan. CREW PRAMs and Decision Trees, \textit{SIAM Journal on Computing} 20(6): 999-1007,
1991.


\bibitem{Nisan94}
Noam Nisan, Mario Szegedy. On the degree of boolean functions as real polynomials.
{\em Computational Complexity}, 4:301-313, 1994


\bibitem{Rubinstein95} D. Rubinstein. Sensitivity vs. Block Sensitivity of Boolean functions, \textit{Combinatorica} 15(2): 297-299, 1995.

\bibitem{Sim} H. U. Simon.
A Tight $\Omega(\log\log n)$-Bound on the Time for Parallel Ram's to Compute Nondegenerated Boolean Functions. {\em FCT'1983}: 439-444.

\bibitem{Virza10} M. Virza. Sensitivity versus block sensitivity of boolean functions. \textit{Information Processing Letters} 111: 433-435, 2011.



\end{thebibliography}
\end{document}